\documentclass[aps,prl,twocolumn,superscriptaddress]{revtex4}
\usepackage{mathrsfs}
\usepackage{epsfig}
\usepackage{graphicx}
\usepackage{amsfonts}
\usepackage[figuresright]{rotating}
\usepackage{amssymb}
\usepackage{amsmath}
\usepackage{dcolumn}
\usepackage{bm}
\usepackage{color}

\def\Re{\rm{Re}}

\def\be{\begin{equation}} \def\ee{\end{equation}}
\def\bea{\begin{eqnarray}} \def\eea{\end{eqnarray}}

\newcommand{\WQCASQC} {Wilczek Quantum Center, School of Physics and Astronomy, Shanghai Jiao Tong University, Shanghai 200240, and Shanghai Research Center for Quantum Sciences, Shanghai 201315, China}

\newcommand{\chongqing} {College of Physics and Electronic Engineering, Chongqing Normal University, Chongqing 401331, China}

\newcommand{\SUSTech}{Key Laboratory of Artificial Structures and Quantum Control, School of Physics and Astronomy, Shanghai Jiao Tong University, Shanghai 200240, China}

\begin{document}
\title{Universal critical dynamics of quantum geometry}

\author{Shuohang Wu}
\thanks{These authors contributed equally to this work.}
\affiliation{\WQCASQC}

\author{Zhengxin Guo}
\thanks{These authors contributed equally to this work.}
\affiliation{\WQCASQC}

\author{Zijian Xiong}
\thanks{These authors contributed equally to this work.}
\affiliation{\chongqing}

\author{Yuan Yao}
\affiliation{\SUSTech}

\author{Zi Cai}
\email{zcai@sjtu.edu.cn}
\affiliation{\WQCASQC}
\affiliation{\SUSTech}

\begin{abstract}  
In this study, we prove that the quantum critical point in the ground state of quantum many-body systems, can also govern the universal dynamical  behavior when the systems are driven far from equilibrium,  which can be captured by the evolution of the quantum geometry of the systems. By investigating  quantum quench dynamics in quadratic fermionic models, we prove that the quantum volume of these systems typically grows linearly over time, with a growth velocity demonstrating universal behavior: its first derivative over the control parameter exhibits a discontinuity at the quantum critical point, with an universal jump value that is independent of specific models, but is crucially determined by the system dimension. This result reveals  universal dynamical properties of non-equilibrium quantum many-body systems.

\end{abstract}


\maketitle

{\it Introduction. ---}
The quantum phase transition, driven by quantum fluctuations rather than by thermal motion,
features non-analytic behavior in the ground state of a quantum many-body system.
Although it is impossible to cool matter to ground state at absolute zero,
the impact of a quantum critical point extends across a broad range of low-energy excited states.
This creates a V-shaped quantum criticality regime that fans out to finite temperatures from the quantum phase transition point~\cite{Sachdev1999}.
Such influence brings the quantum criticality from a concept of theoretical interest to a tangible phenomenon that critically determines the finite-temperature properties of quantum critical materials~\cite{Coleman2005}.
To date,
most studies have concentrated on the thermal equilibrium properties of quantum critical systems,
while the non-equilibrium aspects of the quantum criticality are far less understood~\cite{Torre2010}.

The nonequilibrium physics is generally more complex and richer than equilibrium ones.
Regarding quantum criticality,
substantial efforts have been made to explore the universal dynamical behaviors near quantum critical points.
Examples include universal relaxation at a quantum critical point~\cite{Sachdev1997} and ramping dynamics~\cite{Zurek2005,Dziarmaga2005,Damski2005} governed by the Kibble-Zurek mechanism~\cite{Kibble1976,Zurek1985}.
These phenomena,
despite their non-equilibrium nature,
involve only low-lying excited states above the critical  point, thus are fundamentally tied to the underlying quantum critical points in the ground state.
Compared to these near-equilibrium dynamics, the role of quantum criticality in systems far from the ground state remains less clear.
Notably,
recent progress has shown intrinsic non-equilibrium critical scaling in the quantum quench dynamics of long-range interacting systems~\cite{Titum2020,De2023}.
In this study,
we reveal that even in a system as simple as quadratic free fermions,
the quantum phase transition in the ground state can lead to universal non-analytic behavior in  dynamics far from the ground state,
which can be captured by the quantum geometry~\cite{Resta2011}.

The concept of quantum geometry is introduced to measure distances between quantum states. It contains information of both the phase and amplitude distances, the former is characterized by the Berry curvature,
and the latter by the quantum metric~\cite{Provost1980,Matsuura2010,Ma2010b}.
Compared to the intensively studied Berry curvature in the context of topological physics~\cite{Bohm2003},
the quantum metric has been rarely investigated until very recently.
It has been measured experimentally in various synthetic quantum systems including  cold atoms~\cite{Yi2023},
nitrogen-vacancy centers~\cite{Yu2019}, and superconducting circuits~\cite{Zheng2022}
It is now understood that the full quantum geometry,
including the quantum metric,
influences diverse phenomena\cite{Torma2023}, from flat-band superconductivity and superfluidity~\cite{Peotta2015,Julku2016,Peotta2023,Tian2023,Espinosa2024} to the anomalous Hall effect~\cite{Gianfrate2020,Wang2021,Gao2023},
quantum Fisher information~\cite{Braunstein1994,Zanardi2007,Hauke2016},
electron-phonon interactions~\cite{Yu2023}, and fractional Chern insulators~\cite{Parameswaran2013,Neupert2015,BMera2021,BMera20212}. The quantum geometry has also been investigate in the context of  quantum phase transition\cite{CAROLLO2020}.
In this study,
we demonstrate the crucial role of quantum metric in interpreting nonequilibrium phenomena in quantum systems,
particularly in critical dynamics far from the ground state.

{\it The model ---}
We investigate a general quadratic fermionic model with translational symmetry, whose Hamiltonian in momentum space is expressed as:
\begin{equation}
H=\sum_k C^\dagger_k \hat{\mathcal{H}}_k C_k, \label{eq:Ham}
\end{equation}
where $\hat{\mathcal{H}}_k = \vec{h}_k \cdot \vec{\sigma}$ represents a $2 \times 2$ matrix.
Here,
$\vec{\sigma} = [\hat{\sigma}^X, \hat{\sigma}^Y, \hat{\sigma}^Z]$ denotes the Pauli matrices,
and $\vec{h}_k = [h_k^X, h_k^Y, h_k^Z]$ is a three-dimensional vector.
The annihilation and creation fermionic operators,
$C_k$ and $C^\dagger_k$ respectively,
can be represented either as a two-component fermion $C_k = [c_{\alpha, k}, c_{\beta, k}]^\text{T}$,
with the indices $\alpha, \beta$ signifying the degrees of freedom (such as sublattice, orbital, or spin),
or through the Nambu representation as $C_k = [c_k, c_{-k}^\dagger]^\text{T}$,
which describes a fermionic model with pairing terms.
The energy spectrum of the system is given by $\epsilon_k^{\pm} = \pm \lvert \vec{h}_k \rvert$,
leading to an energy gap defined as $\Delta_k = 2\lvert \vec{h}_k \rvert$.
Typically,
a quantum phase transition in such a quadratic fermionic model is characterized by the closure of this gap and a non-analytic behavior of $\Delta_k$ at one or more specific momenta $k_c$.
When $\Delta_k$ goes to zero,
$\hat{\mathcal{H}}_k$ becomes a zero matrix,
leading to the non-analytic behavior of the wavefunction as long as the singularity of most of the physical quantity.

The non-analytic behavior of $\hat{\mathcal{H}}_k$ at the quantum critical point affects not only static properties such as the ground state energy but also has striking implications for the dynamics far from the ground state.
To elucidate this point,
we examine the quantum quench dynamics, which has been extensively studied in such integral systems~\cite{Barthel2008,Calabrese2011,Mitra2018}.
Different from previous studies, we reexamine this problem from a new perspective of quantum geometry.
The Hamiltonian in Eq.\eqref{eq:Ham} is viewed as a set of independent $k$-modes,
each is a two-level system evolving as a spin precession at a frequency $\Delta_k$.
Therefore, the singularity in $\Delta_k$ critically affects the system's dynamics,
as manifested in the evolution of the quantum geometry.

{\it Quantum metric and quantum state volume ---} Considering two $k$-modes with the initial states $|u^0_k\rangle$ and $|u^0_{k+dk}\rangle$ that are close to each other,
the difference in their precession frequencies will separate these two states after a sufficiently long time.
The distance between the two states can be captured by the Fubini-Study metric $\mathbb{B}(k)$, which is a rank-2 tensor for 2D systems:
\begin{equation}
\mathbb{B}_{ij}(\mathbf{k})=\langle\partial_i \psi_\mathbf{k} |\partial_j \psi_\mathbf{k}\rangle-\langle\partial_i \psi_\mathbf{k} |\psi_\mathbf{k}\rangle \langle \psi_\mathbf{k}|\partial_j \psi_\mathbf{k}\rangle, \label{eq:metric}
\end{equation}
where $\partial_i=\frac{\partial}{\partial k_i}$ with $i=x,y$, and $|\psi_\mathbf{k}\rangle$ is the wavefunction at momentum $\mathbf{k}$.
The imaginary part of $\mathbb{B}_{ij}(k)$ is the well-established Berry curvature,
which characterizes the phase differences of quantum states under a small change of $\mathbf{k}$.
The real part of $\mathbb{B}_{ij}(k)$ is the quantum metric,
which measures the orthogonality of two vectors and is proportional to the amplitude distance of quantum states that are sufficiently close to each other.
To measure the average distance between the quantum states at adjacent momenta,
the quantum state volume (QSV) is introduced\cite{Ozawa20210} to characterize the roughness of the manifold composed of the state vectors in the Brillouin zone (BZ).
The QSV $g$ is defined as
\begin{equation} \label{eq:g_2D}
g_{2D}=\int d\mathbf{k} \sqrt{\det\Re[\mathbb{B}_{ij}(\mathbf{k})]},
\end{equation}
where the integral is over the 1st BZ.
The integrand above is called Riemann volume form once the geometry Re$[\mathbb{B}]$ is given.
It is ``gauge-invariant'' under the coordinate change of $\mathbf{k}$ thereby a genuine physical volume of Hilbert (sub)space.
For a 1D system,
Eq.~\eqref{eq:metric} reduces to a rank-1 tensor (a number) as
\begin{equation}\label{eq:B_1D}
\mathbb{B}(k)= \langle \dot{\psi}_k|\dot{\psi}_k\rangle-\langle \dot{\psi}_k|\psi_k\rangle \langle \psi_k|\dot{\psi}_k \rangle,
\end{equation}
where $\dot{u}_k=\partial u_k/\partial k$, and the QSV $g_{1D}=\int dk \sqrt{\mathbb{B}(k)}$.

\textit{Main result. --- } The key result of this work is outlined as follows: For a translational invariant quadratic fermionic Hamiltonian $\hat{\mathcal{H}}_k(\lambda)$ with a control parameter $\lambda$ as shown in Eq.(\ref{eq:Ham}), let the system initially be prepared in the ground state of $\hat{\mathcal{H}}_k(\lambda_0)$, then evolve following an abrupt parameter change to a final Hamiltonian $\hat{\mathcal{H}}_k(\lambda_f)$. If $\lambda_f$ happens to be close to a gap closing quantum critical point at $\lambda=\lambda_c$, the quantum state volume will exhibit universal dynamical behavior that is independent of most details of systems ({\it e.g.} the specific form of  $\hat{\mathcal{H}}_k(\lambda)$), but only determined by the system dimension.

{\bf Lemma:} \textit{in such a quantum quench dynamics, the quantum state volume typically grows linearly in time (usually accompanied by a periodic oscillation), with a growth velocity $v(\lambda_f)$ depending only on  $\lambda_f$.}

This result can be understood as following: the time dependence of the quantum metric in Eq.(\ref{eq:metric}) and (\ref{eq:B_1D}) comes from the evolution of the wavefunction: $|\psi_k(t)\rangle=e^{i\hat{\mathcal{H}}_k(\lambda_f)t}|\psi_k(0)\rangle$, therefore,  derivatives over k in Eq.(\ref{eq:metric}) and (\ref{eq:B_1D}) lead to terms proportional to $t^2$, which govern the long-time dynamics of $\mathbb{B}(\mathbf{k})$ and the QSV. More rigorously, one can prove that\cite{Suppl}:
\begin{equation}
\det{\Re[\mathbb{B}_{ij}(\mathbf{k})]} = f_\mathbf{k}^0(t)+f_\mathbf{k}^1(t) t+R_\mathbf{k} t^2\sin^2[2 \epsilon_\mathbf{k} t]. \label{eq:Rk}
\end{equation}
The last term dominates the long-time dynamics of $g_{2D}(t)$: a linear growth accompanied by a periodic oscillation. Replacing the oscillating term $|\sin2\epsilon_{\mathbf{k}}t|$ by its average value $\frac 2\pi$ within a period, one can obtain $g_{2D}(t)\sim v(\lambda_f)t$, with a growth velocity $v(\lambda_f)=\frac 2\pi\int d\mathbf{k}\sqrt{R_{\mathbf{k}}}$, and $R_{\mathbf{k}}$ is a time-independent function, which contains the initial state information\cite{Suppl}.

{\bf Theorem:} \textit{the derivative of $v(\lambda_f)$ with respect to $\lambda_f$ exhibits an universal jump at the quantum critical point:
\begin{equation}\label{eq:Cd}
\frac{dv}{d\lambda_f}\bigg |_{\lambda_f=\lambda_c^-} - \frac{dv}{d\lambda_f}\bigg |_{\lambda_f=\lambda_c^+} = \mathcal{N} C_d
\end{equation}
where $\mathcal{N}$ is the number of the gap closing points in the 1st BZ,
and $C_d$ represents a dimensionless universal constant that only depends on the system dimension:
\begin{eqnarray}
C_d = \left\{\begin{array}{cl}
    2\pi, & d=1;   \\
    8, & d=2. \\
\end{array}\right.
\end{eqnarray}
}
{\bf Assumptions:} In the following, we will prove the above universality of $C_d$. The proof is based on two assumptions: (a)the three components of the Hamiltonian vector ${\vec{h}}_{\mathbf{k}}$ are analytic functions in terms of $\mathbf{k}$ and $\lambda$ (even at the quantum critical point); (b)the gap closing points are the only local minimum of the energy spectrum in their infinitesimal vicinity in the BZ.  These two assumptions apply to almost all the realistic models.

{\bf Proof:} For convenience, we first outline the key steps of the proof: Step I: we prove that only those k-modes within a vicinity of the gap closing points $k_c$ in the BZ zone contribute to the non-analytic behavior, which allows for a Taylor expansion of the Hamiltonian vector $\vec{h}_k$ around $k_c$. This is the key step to demonstrate that $C_d$ are independent of the specific form of $\vec{h}_k$.  Step II: various Hamiltonian can be classified into different categories according to the power of the leading order in the Taylor expansion, which simplifies the Hamiltonian vector into a tractable form. Step III: based on these simplified Hamiltonian vectors, we can prove that $C_d$ is indeed universal.

\begin{figure}[tbhp] \centering
\includegraphics[width=0.45\textwidth]{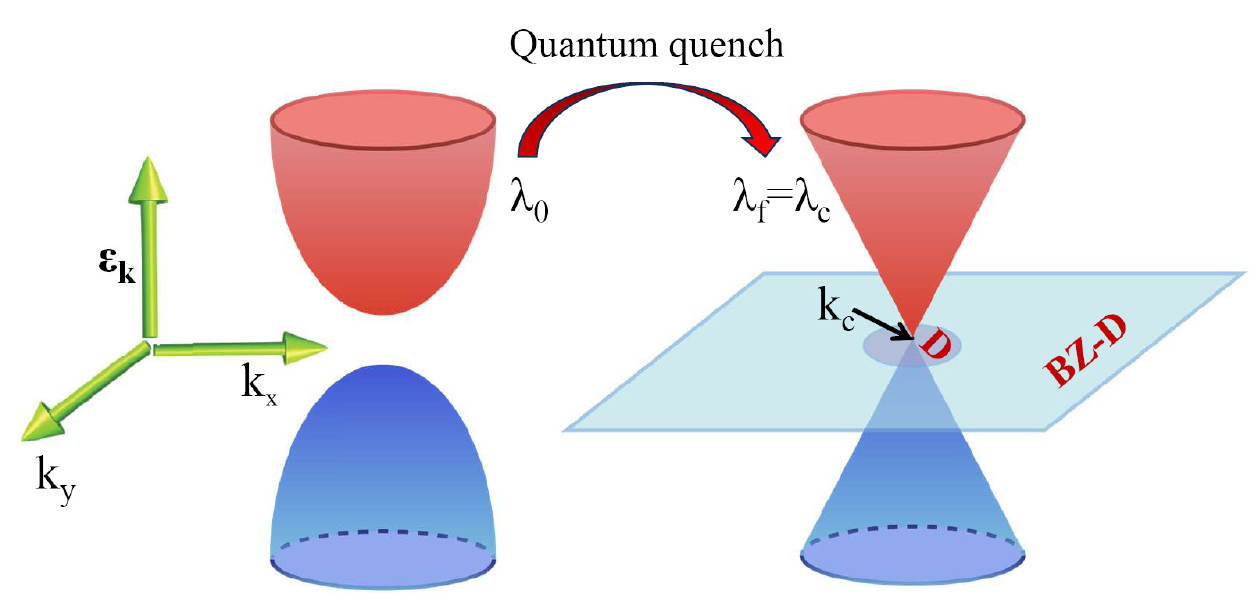}
\caption{Illustration of the quantum quench dynamics in a 2D case.   The shadowed regime  (denoted by $D$) indicates a small vicinity around the gap close point $\mathbf{k}_c$ in the 1st BZ.} \label{Fig:BZ_D}
\end{figure}

Step I:  For simplicity, we take the $\mathcal{N}=1$ case for an example.  There is a single isolated gap closing point ($\mathbf{k}_c$) within the 1st BZ, which can be divided into two regimes: an infinitesimal vicinity around $\mathbf{k}_c$ (denoted by D) and the complement of $D$ (denoted as $\text{BZ}-D$) as shown in Fig.\ref{Fig:BZ_D}. Within the regime $\text{BZ}-D$,
$R_\mathbf{k}$ in Eq.(\ref{eq:Rk}) is an analytical and bounded function for $\mathbf{k} \in \text{BZ}-D$ and $\lambda \in (\lambda_c-\delta \lambda, \lambda_c+\delta \lambda)$ (This is because the $\mathcal{H}_k$ is analytic function in this regime, so is the time-dependent wavefunction $|\psi_\mathbf{k}(t)\rangle$). Therefore, only the $\mathbf{k}$-zeros ($\mathbf{k}$ points satisfying $R_\mathbf{k}(\lambda_c)=0$)  of $R_\mathbf{k}(\lambda_c)$ can contribute the singularity in Eq.(\ref{eq:Cd}) due to the presence of $\sqrt{R_\mathbf{k}(\lambda_c)}$ in the expression of $v_f$. Since $R_\mathbf{k}$ is analytic in $\text{BZ}-D$ and not identically zero, its zeros form a zero-measure subset of the regime $\text{BZ}-D$, thus its contribution to the singularity in Eq.(\ref{eq:Cd}) is neglectable. This explains the fact that $C_d$ is independent of the specific form of $\mathcal{H}_k$, since most k-mode in the BZ do not contribute to it, and only the k-mode within the D regime matters. In addition, one can also understand the jump value in Eq.(\ref{eq:Cd}) is proportional to the number of the gap closing points in the 1st BZ, because the gap closing points are separated, thus their contributions to the singularity are independent with each other.

Step II: Above analysis allows us to focus only on the $\mathbf{k}$-modes within the infinitesimal vicinity around $\mathbf{k}_c$, and the assumption (a) and (b) listed above enable us to perform the Taylor expansion for the Hamiltonian and other related quantities ({\it e.g.} $\epsilon_k^2$)  around $\mathbf{k}_c$. (However, the energy spectrum itself $|\epsilon_k|$ is not analytic at the critical point) {\it For convenience, we assume the gap closing point occurs at $k_c=0$ and $\lambda_f=0$ without loss of generality}. For 1D cases,
$\exists \delta \lambda, \forall \lambda \in (- \delta \lambda, \delta \lambda)$, we can perform the Taylor expansion of $\epsilon_k^2$ in terms of k as
\begin{equation}
\epsilon_k^2 = f(\lambda) k^{2n} + l(\lambda) + o(k^{2n}), \label{eq:epsilon2}
\end{equation}
where the integral $n$ is the lowest order of $k$, and the coefficients satisfy $f(0)$ is a finite value, and $l(0)=0$ since $\epsilon_{k=0}^2=0$ for $\lambda=0$.
The physical implication of the expansion of $\epsilon_k^2$ is that
the form of low-energy effective theory converges as $\lambda\rightarrow\lambda_c$,
which is indeed the case for a large class of lattice models that admit field-theory descriptions. Similarly, one can expand the Hamiltonian vector $\vec{h}_k$ around $k=0$, and one can prove that with proper rotations within the spin space spanned by the Pauli matrices,  the $\vec{h}_k$ after expansion can always be expressed in the following form\cite{Suppl}:
\begin{equation} \label{h_k_reduce_1D}
    \vec{h}_k = [J k^{n} + o(k^{n}), o(k^{n}), \lambda + o(k^{2n})],
\end{equation}
where $J^2=f(0)$, and one can always replace $l(\lambda)$ in Eq.(\ref{eq:epsilon2}) by a new variable $\lambda'$, and the $\lambda$ in Eq.(\ref{h_k_reduce_1D}) satisfy $\lambda^2=\lambda'$.

Similar to the 1D case, with the assumption (a) and (b), we can perform the Taylor expansion for the Hamiltonian vector around $\mathbf{k}=0$ in the 2D case, which can be reduced to the general form\cite{Suppl}:
\begin{small}
\begin{equation} \label{h_k_reduce_2D}
    \vec{h}_k = [J_x k_x^{\alpha} + o(k_x^{\alpha} ,k_y^{\beta}), J_y k_y^{\beta} + o(k_x^{\alpha} ,k_y^{\beta}), \lambda + o(k_x^{2\alpha} ,k_y^{2\beta})].
\end{equation}
\end{small}
where $\alpha$ and $\beta$ are integers that represent the lowest orders appear in the Taylor expansion.

Step III: Using the simplified forms of the $\vec{h}_k$ in Eq.(\ref{h_k_reduce_1D}) and (\ref{h_k_reduce_2D}) and ignoring the higher order terms (since we only focus on an infinitesimal regime around $k=0$), we can derive the analytic form of the growth velocity as well as  the universal constant $C_d$. For 1D cases, the growth rate of the quantum state volume is given by~\cite{Suppl}
\begin{equation}
    v_\text{1D}(\lambda_f)= 2\left(\Lambda^nJ\!-\!\lambda_f\arctan{\frac{\Lambda^nJ}{\lambda_f}}\right), \label{eq:vf1D}
\end{equation}
where $\Lambda$ is the radius of the D regime. By substitute Eq.(\ref{eq:vf1D}) into Eq.(\ref{eq:Cd}), one can easily obtain that $C_1=2\pi$, which is independent of all the parameters in the original or reduce Hamiltonian vectors. Similarly, for the 2D case, we can perform the integral over $\mathbf{k}$ and obtain~\cite{Suppl}
\begin{equation}\small
    v_\text{2D}(\lambda_f) 
    = \frac{4(\sqrt{\lambda_f^2+\Lambda^2}-|\lambda_f|)^2}{\sqrt{\lambda_f^2+\Lambda^2}},
\end{equation}
which leads to the universal constant $C_2=8$.

\begin{figure*}[tbhp] \centering
\includegraphics[width=1\textwidth]{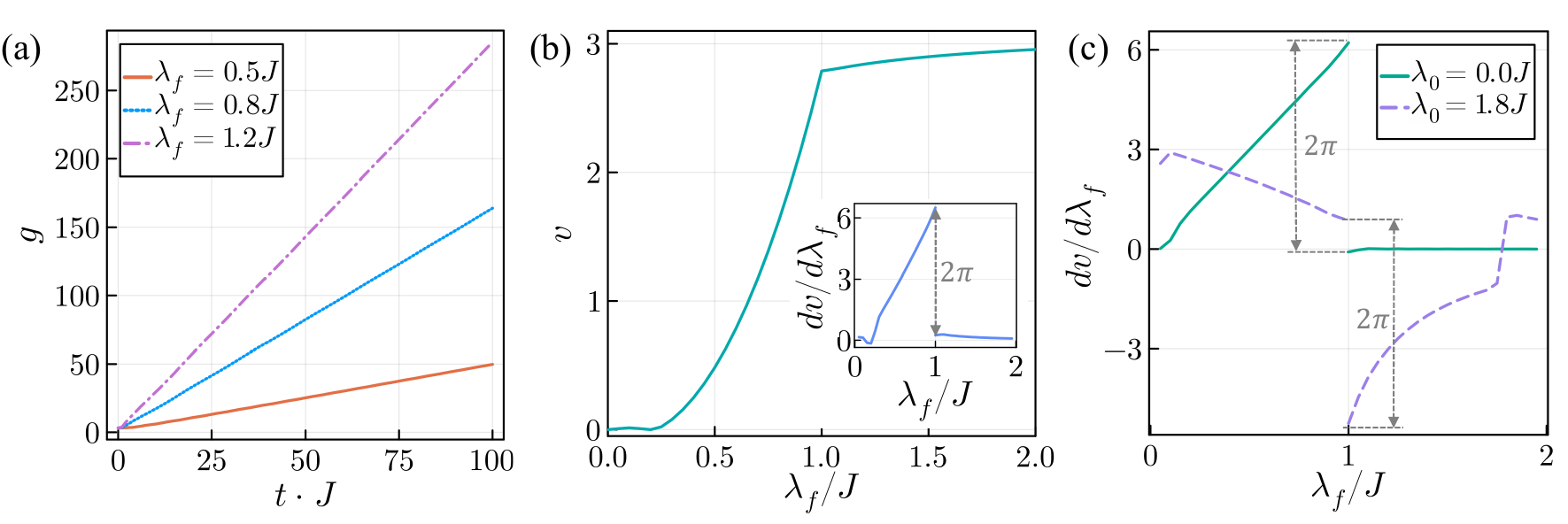}
\caption{(Color online) (a) Quantum state volume $g(t)$ for the 1D Kitaev model in the quantum quench dynamics starting from the same initial states with $\lambda_0=0.2J$, but evolving under the Hamiltonian with different $\lambda_f$.
(b) The growth velocity $v$ as a function of $\lambda_f$. The inset is its derivative with respect to $\lambda_f$ , which exhibits a discontinuity with an universal jump value $2\pi$.
(c) $dv/d\lambda_f$ as a function of $\lambda_f$ starting from initial states with different $\lambda_0$, which share the same jump value of $2\pi$ at $\lambda_f=J$.}\label{fig:fig1}
\end{figure*}

\begin{figure*}[tbhp] \centering
\includegraphics[width=1\textwidth]{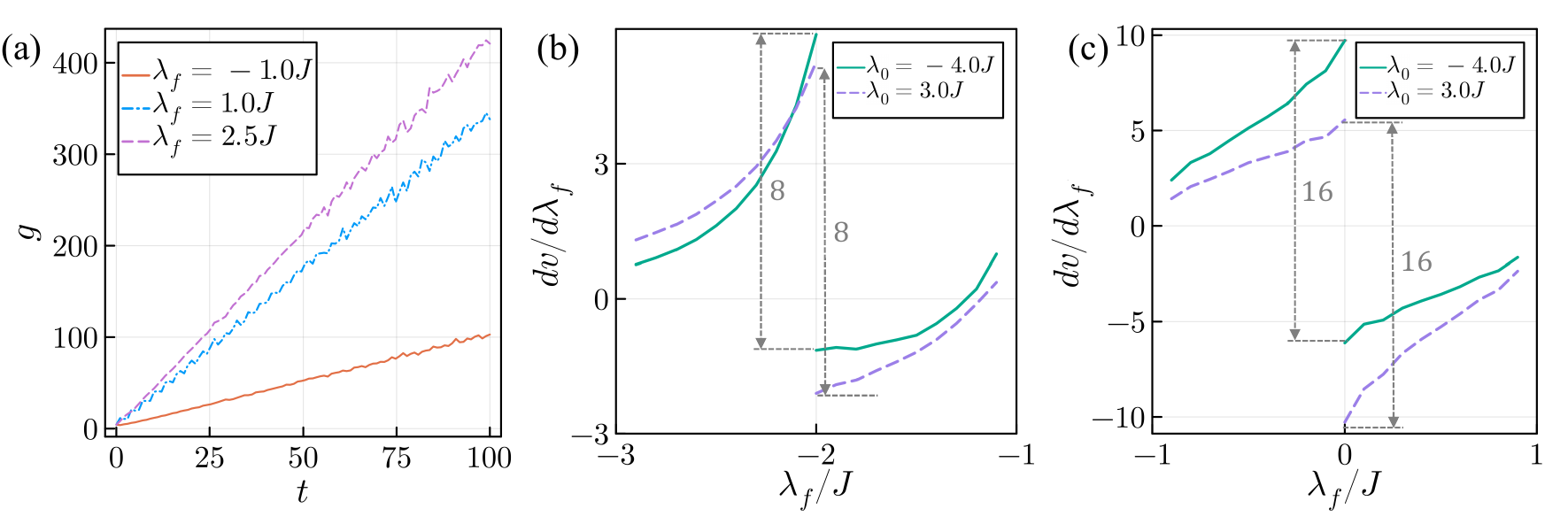}
\caption{(Color online) (a) Quantum volume $g(t)$ for the 2D Qi-Wu-Zhang model in the quantum quench dynamics starting from the same initial states with $\lambda_0=-0.5J$, but evolving under the Hamiltonian with different $\lambda_f$.
(b) and (c) $dv/d\lambda_f$ as a function of $\lambda_f$ around the different quantum critical points at (b) $\lambda_f=-2J$ and (c) $\lambda_f=0$.  The number of the gap closing points in the 1st BZ is $\mathcal{N}=1$ ($\mathcal{N}=2$) for the critical point at $\lambda_f=-2J$ ($\lambda_f=0$).}\label{Zhangqvol}
\end{figure*}

{\bf Example I:} In spite of the exact proof listed above, it is helpful to numerically verify the analytic results in specific examples. To this end, we consider well-studied models in 1D and 2D.  The first example we consider is the 1D Kitaev chain~\cite{Kitaev2001}, characterized by a Hamiltonian represented by Eq.~\eqref{eq:Ham} in momentum space. Here, $C_k = [c_k, c_{-k}^\dag]$ denotes the spinless fermionic operator in the Nambu representation, and the Hamiltonian vector is given by~\cite{Kitaev2001}
\begin{equation}
\vec{h}(k) = [0, i\Delta\sin k, J\cos k + \lambda(t)],
\end{equation}
where $J$ and $\Delta$ represent the hopping and pairing amplitudes of fermions on adjacent sites, respectively.
For simplicity, we assume $\Delta = J$.
The term $\lambda(t)$,
signifying the chemical potential,
acts as a time-dependent control parameter in this context.
Notably, this model facilitates a topological quantum phase transition in its ground state at the critical threshold $\lambda_c = J$,
where the gap closes at $k_c = \pi$. Near the critical point,
the energy spectrum approximates to $\epsilon_k \sim J|k - k_c|$.
We examine the quench dynamics of the model,
starting from the ground state of the Hamiltonian with a control parameter $\lambda=\lambda_0$.
This parameter is abruptly altered to $\lambda_f$,
and we study the subsequent evolution under the new Hamiltonian.
The dynamics of $g_{1D}(t)$,
originating from the same initial state but evolving under various $\lambda_f$,
is depicted in Fig.~\ref{fig:fig1}(a),
revealing a linear increase in $g(t)$.
The velocity of growth $v$,
as a function of $\lambda_f$,
is presented in Fig.~\ref{fig:fig1}(b),
displaying a distinct cusp at the quantum critical point $\lambda_f=J$.
This cusp signifies a discontinuity in the first-order derivative of $v(\lambda_f)$ at $\lambda_f=\lambda_c$,
as illustrated in the inset of Fig.~\ref{fig:fig1}(b).
Noteworthy is Fig.~\ref{fig:fig1}(c),
which indicates that the magnitude of the discontinuity $C_1=2\pi$ is independent of the initial state parameter $\lambda_0$. Other 1D non-interacting models with a quantum phase transition, such as the Su-Schrieffer-Heeger model\cite{Su1979}, 1D spinless fermion model with staggered chemical potential, have also been examined numerically\cite{Suppl}, and all of them satisfy the relation in Eq.(\ref{eq:Cd}).

{\bf Example II:} For 2D system, we consider the Qi-Wu-Zhang (QWZ) model~\cite{Qi2006}, whose Hamiltonian vector is defined as:
\begin{equation}\small
\vec{h}_{QWZ}(\mathbf{k}) = [J\sin k_x, J\sin k_y, J(\cos k_x + \cos k_y) + \lambda(t)],\label{eq:QWZ}
\end{equation}
where $(k_x, k_y) \in (-\pi, \pi]$.
This model exhibits topological quantum phase transitions at $\lambda = 0, \pm2J$ separately.
In the case of $\lambda = 0$, there are two gap closing points in the 1st BZ ($\mathcal{N}=2$):
$\mathbf{k}_c^1=(0,\pi)$ and $\mathbf{k}_c^2=(\pi,0)$; For the cases with $\lambda = 2J(-2J)$,
there is only one gap closing point ($\mathcal{N}=1$) at $\mathbf{k}_c=(0,0)$ ($\mathbf{k}_c=(\pi,\pi)$).
We initialize the system in the ground state of the Hamiltonian.~(\ref{eq:QWZ}) with $\lambda = \lambda_0$,
then abruptly change the parameter to $\lambda_f$ and examine the subsequent evolution of the quantum volume $g(t)$.
$g(t)$ demonstrates linear growth over time,
as depicted in Fig.~\ref{Zhangqvol}(a).
From this, we derive the growth velocity $v(\lambda_f)$ and evaluate its derivative with respect to $\lambda_f$.
Fig.~\ref{Zhangqvol}(b) and (c) indicate that the universal jump values at $\lambda_f=-2J$ and $\lambda_f=0$ satisfy Eq.~\eqref{eq:Cd} with $C_2 = 8$ and $\mathcal{N}=1$ and  $\mathcal{N}=2$ respectively. Other 2D models ({\it e.g} a spinless fermioin in a Boron-Nitride-like lattice\cite{Cai2008}) have also been examined numerically\cite{Suppl}.


 {\it Conclusion and outlook. ---} In conclusion, we demonstrate that quantum criticality can govern the non-equilibrium dynamics far from the ground state and induce universal dynamical behavior.
Although the quantum volume relies on the non-interacting band structure,
its universal velocity jump is completely determined by the $\mathbf{k}$-asymptotic behavior \textit{locally} around the gap closing points.
Thus,
we expect a potential extension of such volume to interacting systems\cite{Souza2000} admitting an effective field theory with emergent translation symmetry around criticality,
such as Dirac physics and other conformal field theories~\cite{Francesco:2012aa,Wehling:2014aa},
which is a generic feature in low dimension~\cite{Gogolin:2004aa}.

{\it Acknowledgments}.---  This work is supported by the National Key Research and Development Program of China (Grant No. 2020YFA0309000), NSFC of  China (Grant No.12174251), Natural Science Foundation of Shanghai (Grant No.22ZR142830),  Shanghai Municipal Science and Technology Major Project (Grant No.2019SHZDZX01). Z.X. is supported by the start-up funding of CQNU (Grant No.24XLB010)


\begin{thebibliography}{50}
\expandafter\ifx\csname natexlab\endcsname\relax\def\natexlab#1{#1}\fi
\expandafter\ifx\csname bibnamefont\endcsname\relax
  \def\bibnamefont#1{#1}\fi
\expandafter\ifx\csname bibfnamefont\endcsname\relax
  \def\bibfnamefont#1{#1}\fi
\expandafter\ifx\csname citenamefont\endcsname\relax
  \def\citenamefont#1{#1}\fi
\expandafter\ifx\csname url\endcsname\relax
  \def\url#1{\texttt{#1}}\fi
\expandafter\ifx\csname urlprefix\endcsname\relax\def\urlprefix{URL }\fi
\providecommand{\bibinfo}[2]{#2}
\providecommand{\eprint}[2][]{\url{#2}}

\bibitem[{\citenamefont{Sachdev}(1999)}]{Sachdev1999}
\bibinfo{author}{\bibfnamefont{S.}~\bibnamefont{Sachdev}},
  \emph{\bibinfo{title}{Quantum Phase Transitions}}
  (\bibinfo{publisher}{~Cambridge University Press, Cambridge},
  \bibinfo{year}{1999}).

\bibitem[{\citenamefont{Coleman and Schofield}(2005)}]{Coleman2005}
\bibinfo{author}{\bibfnamefont{P.}~\bibnamefont{Coleman}} \bibnamefont{and}
  \bibinfo{author}{\bibfnamefont{A.~J.} \bibnamefont{Schofield}},
  \bibinfo{journal}{Nature} \textbf{\bibinfo{volume}{433}},
  \bibinfo{pages}{226} (\bibinfo{year}{2005}).

\bibitem[{\citenamefont{Dalla~Torre et~al.}(2010)\citenamefont{Dalla~Torre,
  Demler, Giamarchi, and Altman}}]{Torre2010}
\bibinfo{author}{\bibfnamefont{E.~G.} \bibnamefont{Dalla~Torre}},
  \bibinfo{author}{\bibfnamefont{E.}~\bibnamefont{Demler}},
  \bibinfo{author}{\bibfnamefont{T.}~\bibnamefont{Giamarchi}},
  \bibnamefont{and} \bibinfo{author}{\bibfnamefont{E.}~\bibnamefont{Altman}},
  \bibinfo{journal}{Nat. Phys.} \textbf{\bibinfo{volume}{6}},
  \bibinfo{pages}{806} (\bibinfo{year}{2010}).

\bibitem[{\citenamefont{Sachdev and Young}(1997)}]{Sachdev1997}
\bibinfo{author}{\bibfnamefont{S.}~\bibnamefont{Sachdev}} \bibnamefont{and}
  \bibinfo{author}{\bibfnamefont{A.~P.} \bibnamefont{Young}},
  \bibinfo{journal}{Phys. Rev. Lett.} \textbf{\bibinfo{volume}{78}},
  \bibinfo{pages}{2220} (\bibinfo{year}{1997}).

\bibitem[{\citenamefont{Zurek et~al.}(2005)\citenamefont{Zurek, Dorner, and
  Zoller}}]{Zurek2005}
\bibinfo{author}{\bibfnamefont{W.~H.} \bibnamefont{Zurek}},
  \bibinfo{author}{\bibfnamefont{U.}~\bibnamefont{Dorner}}, \bibnamefont{and}
  \bibinfo{author}{\bibfnamefont{P.}~\bibnamefont{Zoller}},
  \bibinfo{journal}{Phys. Rev. Lett.} \textbf{\bibinfo{volume}{95}},
  \bibinfo{pages}{105701} (\bibinfo{year}{2005}).

\bibitem[{\citenamefont{Dziarmaga}(2005)}]{Dziarmaga2005}
\bibinfo{author}{\bibfnamefont{J.}~\bibnamefont{Dziarmaga}},
  \bibinfo{journal}{Phys. Rev. Lett.} \textbf{\bibinfo{volume}{95}},
  \bibinfo{pages}{245701} (\bibinfo{year}{2005}).

\bibitem[{\citenamefont{Damski}(2005)}]{Damski2005}
\bibinfo{author}{\bibfnamefont{B.}~\bibnamefont{Damski}},
  \bibinfo{journal}{Phys. Rev. Lett.} \textbf{\bibinfo{volume}{95}},
  \bibinfo{pages}{035701} (\bibinfo{year}{2005}).

\bibitem[{\citenamefont{Kibble}(1976)}]{Kibble1976}
\bibinfo{author}{\bibfnamefont{T.~W.~B.} \bibnamefont{Kibble}},
  \bibinfo{journal}{Journal of Physics A: Mathematical and General}
  \textbf{\bibinfo{volume}{9}}, \bibinfo{pages}{1387} (\bibinfo{year}{1976}).

\bibitem[{\citenamefont{Zurek}(1985)}]{Zurek1985}
\bibinfo{author}{\bibfnamefont{W.}~\bibnamefont{Zurek}},
  \bibinfo{journal}{Nature} \textbf{\bibinfo{volume}{317}},
  \bibinfo{pages}{505} (\bibinfo{year}{1985}).

\bibitem[{\citenamefont{Titum and Maghrebi}(2020)}]{Titum2020}
\bibinfo{author}{\bibfnamefont{P.}~\bibnamefont{Titum}} \bibnamefont{and}
  \bibinfo{author}{\bibfnamefont{M.~F.} \bibnamefont{Maghrebi}},
  \bibinfo{journal}{Phys. Rev. Lett.} \textbf{\bibinfo{volume}{125}},
  \bibinfo{pages}{040602} (\bibinfo{year}{2020}).

\bibitem[{\citenamefont{{De} et~al.}(2023)\citenamefont{{De}, {Cook},
  {Collins}, {Morong}, {Paz}, {Titum}, {Pagano}, {Gorshkov}, {Maghrebi}, and
  {Monroe}}}]{De2023}
\bibinfo{author}{\bibfnamefont{A.}~\bibnamefont{{De}}},
  \bibinfo{author}{\bibfnamefont{P.}~\bibnamefont{{Cook}}},
  \bibinfo{author}{\bibfnamefont{K.}~\bibnamefont{{Collins}}},
  \bibinfo{author}{\bibfnamefont{W.}~\bibnamefont{{Morong}}},
  \bibinfo{author}{\bibfnamefont{D.}~\bibnamefont{{Paz}}},
  \bibinfo{author}{\bibfnamefont{P.}~\bibnamefont{{Titum}}},
  \bibinfo{author}{\bibfnamefont{G.}~\bibnamefont{{Pagano}}},
  \bibinfo{author}{\bibfnamefont{A.~V.} \bibnamefont{{Gorshkov}}},
  \bibinfo{author}{\bibfnamefont{M.}~\bibnamefont{{Maghrebi}}},
  \bibnamefont{and} \bibinfo{author}{\bibfnamefont{C.}~\bibnamefont{{Monroe}}},
  \bibinfo{journal}{arXiv e-prints} \bibinfo{eid}{arXiv:2309.10856}
  (\bibinfo{year}{2023}), \eprint{2309.10856}.

\bibitem[{\citenamefont{Resta}(2011)}]{Resta2011}
\bibinfo{author}{\bibfnamefont{R.}~\bibnamefont{Resta}}, \bibinfo{journal}{Eur.
  Phys. J. B} \textbf{\bibinfo{volume}{79}}, \bibinfo{pages}{121}
  (\bibinfo{year}{2011}).

\bibitem[{\citenamefont{Provost and G.Vallee}(1980)}]{Provost1980}
\bibinfo{author}{\bibfnamefont{J.~P.} \bibnamefont{Provost}} \bibnamefont{and}
  \bibinfo{author}{\bibnamefont{G.Vallee}}, \bibinfo{journal}{Commun. Math.
  Phys.} \textbf{\bibinfo{volume}{76}}, \bibinfo{pages}{289}
  (\bibinfo{year}{1980}).

\bibitem[{\citenamefont{Matsuura and Ryu}(2010)}]{Matsuura2010}
\bibinfo{author}{\bibfnamefont{S.}~\bibnamefont{Matsuura}} \bibnamefont{and}
  \bibinfo{author}{\bibfnamefont{S.}~\bibnamefont{Ryu}},
  \bibinfo{journal}{Phys. Rev. B} \textbf{\bibinfo{volume}{82}},
  \bibinfo{pages}{245113} (\bibinfo{year}{2010}).

\bibitem[{\citenamefont{Ma et~al.}(2010)\citenamefont{Ma, Chen, Fan, and
  Liu}}]{Ma2010b}
\bibinfo{author}{\bibfnamefont{Y.-Q.} \bibnamefont{Ma}},
  \bibinfo{author}{\bibfnamefont{S.}~\bibnamefont{Chen}},
  \bibinfo{author}{\bibfnamefont{H.}~\bibnamefont{Fan}}, \bibnamefont{and}
  \bibinfo{author}{\bibfnamefont{W.-M.} \bibnamefont{Liu}},
  \bibinfo{journal}{Phys. Rev. B} \textbf{\bibinfo{volume}{81}},
  \bibinfo{pages}{245129} (\bibinfo{year}{2010}).

\bibitem[{\citenamefont{Bohm et~al.}(2003)\citenamefont{Bohm, Mostafazadeh, Ko,
  Niu, and Zwanziger}}]{Bohm2003}
\bibinfo{author}{\bibfnamefont{A.}~\bibnamefont{Bohm}},
  \bibinfo{author}{\bibfnamefont{A.}~\bibnamefont{Mostafazadeh}},
  \bibinfo{author}{\bibfnamefont{H.}~\bibnamefont{Ko}},
  \bibinfo{author}{\bibfnamefont{Q.}~\bibnamefont{Niu}}, \bibnamefont{and}
  \bibinfo{author}{\bibfnamefont{J.}~\bibnamefont{Zwanziger}},
  \emph{\bibinfo{title}{The Geometric Phase in Quantum Systems}}
  (\bibinfo{publisher}{~Springer, Berlin}, \bibinfo{year}{2003}).

\bibitem[{\citenamefont{Yi et~al.}(2023)\citenamefont{Yi, Yu, Yuan, Jiao, Yang,
  Jiang, Zhang, Chen, and Pan}}]{Yi2023}
\bibinfo{author}{\bibfnamefont{C.-R.} \bibnamefont{Yi}},
  \bibinfo{author}{\bibfnamefont{J.}~\bibnamefont{Yu}},
  \bibinfo{author}{\bibfnamefont{H.}~\bibnamefont{Yuan}},
  \bibinfo{author}{\bibfnamefont{R.-H.} \bibnamefont{Jiao}},
  \bibinfo{author}{\bibfnamefont{Y.-M.} \bibnamefont{Yang}},
  \bibinfo{author}{\bibfnamefont{X.}~\bibnamefont{Jiang}},
  \bibinfo{author}{\bibfnamefont{J.-Y.} \bibnamefont{Zhang}},
  \bibinfo{author}{\bibfnamefont{S.}~\bibnamefont{Chen}}, \bibnamefont{and}
  \bibinfo{author}{\bibfnamefont{J.-W.} \bibnamefont{Pan}},
  \bibinfo{journal}{Phys. Rev. Res.} \textbf{\bibinfo{volume}{5}},
  \bibinfo{pages}{L032016} (\bibinfo{year}{2023}).

\bibitem[{\citenamefont{Yu et~al.}(2019)\citenamefont{Yu, Yang, Gong, Cao, Lu,
  Liu, Zhang, Plenio, Jelezko, Ozawa et~al.}}]{Yu2019}
\bibinfo{author}{\bibfnamefont{M.}~\bibnamefont{Yu}},
  \bibinfo{author}{\bibfnamefont{P.}~\bibnamefont{Yang}},
  \bibinfo{author}{\bibfnamefont{M.}~\bibnamefont{Gong}},
  \bibinfo{author}{\bibfnamefont{Q.}~\bibnamefont{Cao}},
  \bibinfo{author}{\bibfnamefont{Q.}~\bibnamefont{Lu}},
  \bibinfo{author}{\bibfnamefont{H.}~\bibnamefont{Liu}},
  \bibinfo{author}{\bibfnamefont{S.}~\bibnamefont{Zhang}},
  \bibinfo{author}{\bibfnamefont{M.~B.} \bibnamefont{Plenio}},
  \bibinfo{author}{\bibfnamefont{F.}~\bibnamefont{Jelezko}},
  \bibinfo{author}{\bibfnamefont{T.}~\bibnamefont{Ozawa}},
  \bibnamefont{et~al.}, \bibinfo{journal}{Natl. Sci. Rev.}
  \textbf{\bibinfo{volume}{7}}, \bibinfo{pages}{254} (\bibinfo{year}{2019}).

\bibitem[{\citenamefont{Zheng et~al.}(2022)\citenamefont{Zheng, Xu, Ma, Li,
  Dong, Zhang, X.Wang, Sun, Wu, Zhao et~al.}}]{Zheng2022}
\bibinfo{author}{\bibfnamefont{W.}~\bibnamefont{Zheng}},
  \bibinfo{author}{\bibfnamefont{J.}~\bibnamefont{Xu}},
  \bibinfo{author}{\bibfnamefont{Z.}~\bibnamefont{Ma}},
  \bibinfo{author}{\bibfnamefont{Y.}~\bibnamefont{Li}},
  \bibinfo{author}{\bibfnamefont{Y.}~\bibnamefont{Dong}},
  \bibinfo{author}{\bibfnamefont{Y.}~\bibnamefont{Zhang}},
  \bibinfo{author}{\bibnamefont{X.Wang}},
  \bibinfo{author}{\bibfnamefont{G.}~\bibnamefont{Sun}},
  \bibinfo{author}{\bibfnamefont{P.}~\bibnamefont{Wu}},
  \bibinfo{author}{\bibfnamefont{J.}~\bibnamefont{Zhao}}, \bibnamefont{et~al.},
  \bibinfo{journal}{Chin. Phys. Lett.} \textbf{\bibinfo{volume}{39}},
  \bibinfo{pages}{100202} (\bibinfo{year}{2022}).

\bibitem[{\citenamefont{T\"orm\"a}(2023)}]{Torma2023}
\bibinfo{author}{\bibfnamefont{P.}~\bibnamefont{T\"orm\"a}},
  \bibinfo{journal}{Phys. Rev. Lett.} \textbf{\bibinfo{volume}{131}},
  \bibinfo{pages}{240001} (\bibinfo{year}{2023}).

\bibitem[{\citenamefont{{Peotta} and {T{\"o}rm{\"a}}}(2015)}]{Peotta2015}
\bibinfo{author}{\bibfnamefont{S.}~\bibnamefont{{Peotta}}} \bibnamefont{and}
  \bibinfo{author}{\bibfnamefont{P.}~\bibnamefont{{T{\"o}rm{\"a}}}},
  \bibinfo{journal}{Nat. Comms} \textbf{\bibinfo{volume}{6}},
  \bibinfo{pages}{8944} (\bibinfo{year}{2015}).

\bibitem[{\citenamefont{Julku et~al.}(2016)\citenamefont{Julku, Peotta,
  Vanhala, Kim, and T\"orm\"a}}]{Julku2016}
\bibinfo{author}{\bibfnamefont{A.}~\bibnamefont{Julku}},
  \bibinfo{author}{\bibfnamefont{S.}~\bibnamefont{Peotta}},
  \bibinfo{author}{\bibfnamefont{T.~I.} \bibnamefont{Vanhala}},
  \bibinfo{author}{\bibfnamefont{D.-H.} \bibnamefont{Kim}}, \bibnamefont{and}
  \bibinfo{author}{\bibfnamefont{P.}~\bibnamefont{T\"orm\"a}},
  \bibinfo{journal}{Phys. Rev. Lett.} \textbf{\bibinfo{volume}{117}},
  \bibinfo{pages}{045303} (\bibinfo{year}{2016}).

\bibitem[{\citenamefont{{Peotta} et~al.}(2023)\citenamefont{{Peotta},
  {Huhtinen}, and {T{\"o}rm{\"a}}}}]{Peotta2023}
\bibinfo{author}{\bibfnamefont{S.}~\bibnamefont{{Peotta}}},
  \bibinfo{author}{\bibfnamefont{K.-E.} \bibnamefont{{Huhtinen}}},
  \bibnamefont{and}
  \bibinfo{author}{\bibfnamefont{P.}~\bibnamefont{{T{\"o}rm{\"a}}}},
  \bibinfo{journal}{arXiv e-prints} \bibinfo{eid}{arXiv:2308.08248}
  (\bibinfo{year}{2023}), \eprint{2308.08248}.

\bibitem[{\citenamefont{Tian et~al.}(2023)\citenamefont{Tian, Gao, abd Shi~Che,
  Xu, Cheung, Watanabe, Taniguchi, Randeria, Zhang, Lau et~al.}}]{Tian2023}
\bibinfo{author}{\bibfnamefont{H.}~\bibnamefont{Tian}},
  \bibinfo{author}{\bibfnamefont{X.}~\bibnamefont{Gao}},
  \bibinfo{author}{\bibfnamefont{Y.~Z.} \bibnamefont{abd Shi~Che}},
  \bibinfo{author}{\bibfnamefont{T.}~\bibnamefont{Xu}},
  \bibinfo{author}{\bibfnamefont{P.}~\bibnamefont{Cheung}},
  \bibinfo{author}{\bibfnamefont{K.}~\bibnamefont{Watanabe}},
  \bibinfo{author}{\bibfnamefont{T.}~\bibnamefont{Taniguchi}},
  \bibinfo{author}{\bibfnamefont{M.}~\bibnamefont{Randeria}},
  \bibinfo{author}{\bibfnamefont{F.}~\bibnamefont{Zhang}},
  \bibinfo{author}{\bibfnamefont{C.~N.} \bibnamefont{Lau}},
  \bibnamefont{et~al.}, \bibinfo{journal}{Nature}
  \textbf{\bibinfo{volume}{641}}, \bibinfo{pages}{440} (\bibinfo{year}{2023}).

\bibitem[{\citenamefont{de~Jesús Espinosa-Champo and
  Naumis}(2023)}]{Espinosa2024}
\bibinfo{author}{\bibfnamefont{A.}~\bibnamefont{de~Jesús Espinosa-Champo}}
  \bibnamefont{and} \bibinfo{author}{\bibfnamefont{G.~G.}
  \bibnamefont{Naumis}}, \bibinfo{journal}{Journal of Physics: Condensed
  Matter} \textbf{\bibinfo{volume}{36}}, \bibinfo{pages}{015502}
  (\bibinfo{year}{2023}).

\bibitem[{\citenamefont{Gianfrate et~al.}(2020)\citenamefont{Gianfrate, Bleu,
  Dominici, Ardizzone, Giorgi, Ballarini, Lerario, West, Pfeiffer, Solnyshkov
  et~al.}}]{Gianfrate2020}
\bibinfo{author}{\bibfnamefont{A.}~\bibnamefont{Gianfrate}},
  \bibinfo{author}{\bibfnamefont{O.}~\bibnamefont{Bleu}},
  \bibinfo{author}{\bibfnamefont{L.}~\bibnamefont{Dominici}},
  \bibinfo{author}{\bibfnamefont{V.}~\bibnamefont{Ardizzone}},
  \bibinfo{author}{\bibfnamefont{M.~D.} \bibnamefont{Giorgi}},
  \bibinfo{author}{\bibfnamefont{D.}~\bibnamefont{Ballarini}},
  \bibinfo{author}{\bibfnamefont{G.}~\bibnamefont{Lerario}},
  \bibinfo{author}{\bibfnamefont{K.}~\bibnamefont{West}},
  \bibinfo{author}{\bibfnamefont{L.~N.} \bibnamefont{Pfeiffer}},
  \bibinfo{author}{\bibfnamefont{D.~D.} \bibnamefont{Solnyshkov}},
  \bibnamefont{et~al.}, \bibinfo{journal}{Nature}
  \textbf{\bibinfo{volume}{578}}, \bibinfo{pages}{381} (\bibinfo{year}{2020}).

\bibitem[{\citenamefont{Wang et~al.}(2021)\citenamefont{Wang, Gao, and
  Xiao}}]{Wang2021}
\bibinfo{author}{\bibfnamefont{C.}~\bibnamefont{Wang}},
  \bibinfo{author}{\bibfnamefont{Y.}~\bibnamefont{Gao}}, \bibnamefont{and}
  \bibinfo{author}{\bibfnamefont{D.}~\bibnamefont{Xiao}},
  \bibinfo{journal}{Phys. Rev. Lett.} \textbf{\bibinfo{volume}{127}},
  \bibinfo{pages}{277201} (\bibinfo{year}{2021}).

\bibitem[{\citenamefont{Gao et~al.}(2023)\citenamefont{Gao, Liu, Qiu, Ghosh,
  Trevisan, Onishi, Hu, Qian, Tien, Chen et~al.}}]{Gao2023}
\bibinfo{author}{\bibfnamefont{A.}~\bibnamefont{Gao}},
  \bibinfo{author}{\bibfnamefont{Y.-F.} \bibnamefont{Liu}},
  \bibinfo{author}{\bibfnamefont{J.-X.} \bibnamefont{Qiu}},
  \bibinfo{author}{\bibfnamefont{B.}~\bibnamefont{Ghosh}},
  \bibinfo{author}{\bibfnamefont{T.~V.} \bibnamefont{Trevisan}},
  \bibinfo{author}{\bibfnamefont{Y.}~\bibnamefont{Onishi}},
  \bibinfo{author}{\bibfnamefont{C.}~\bibnamefont{Hu}},
  \bibinfo{author}{\bibfnamefont{T.}~\bibnamefont{Qian}},
  \bibinfo{author}{\bibfnamefont{H.-J.} \bibnamefont{Tien}},
  \bibinfo{author}{\bibfnamefont{S.-W.} \bibnamefont{Chen}},
  \bibnamefont{et~al.}, \bibinfo{journal}{Science}
  \textbf{\bibinfo{volume}{381}}, \bibinfo{pages}{181} (\bibinfo{year}{2023}).

\bibitem[{\citenamefont{Braunstein and Caves}(1994)}]{Braunstein1994}
\bibinfo{author}{\bibfnamefont{S.~L.} \bibnamefont{Braunstein}}
  \bibnamefont{and} \bibinfo{author}{\bibfnamefont{C.~M.} \bibnamefont{Caves}},
  \bibinfo{journal}{Phys. Rev. Lett.} \textbf{\bibinfo{volume}{72}},
  \bibinfo{pages}{3439} (\bibinfo{year}{1994}).

\bibitem[{\citenamefont{Zanardi et~al.}(2007)\citenamefont{Zanardi, Giorda, and
  Cozzini}}]{Zanardi2007}
\bibinfo{author}{\bibfnamefont{P.}~\bibnamefont{Zanardi}},
  \bibinfo{author}{\bibfnamefont{P.}~\bibnamefont{Giorda}}, \bibnamefont{and}
  \bibinfo{author}{\bibfnamefont{M.}~\bibnamefont{Cozzini}},
  \bibinfo{journal}{Phys. Rev. Lett.} \textbf{\bibinfo{volume}{99}},
  \bibinfo{pages}{100603} (\bibinfo{year}{2007}).

\bibitem[{\citenamefont{Hauke et~al.}(2016)\citenamefont{Hauke, Heyl,
  Tagliacozzo, and Zoller}}]{Hauke2016}
\bibinfo{author}{\bibfnamefont{P.}~\bibnamefont{Hauke}},
  \bibinfo{author}{\bibfnamefont{M.}~\bibnamefont{Heyl}},
  \bibinfo{author}{\bibfnamefont{L.}~\bibnamefont{Tagliacozzo}},
  \bibnamefont{and} \bibinfo{author}{\bibfnamefont{P.}~\bibnamefont{Zoller}},
  \bibinfo{journal}{Nature Physics} \textbf{\bibinfo{volume}{12}},
  \bibinfo{pages}{778} (\bibinfo{year}{2016}).

\bibitem[{\citenamefont{{Yu} et~al.}(2023)\citenamefont{{Yu}, {Ciccarino},
  {Bianco}, {Errea}, {Narang}, and {Bernevig}}}]{Yu2023}
\bibinfo{author}{\bibfnamefont{J.}~\bibnamefont{{Yu}}},
  \bibinfo{author}{\bibfnamefont{C.~J.} \bibnamefont{{Ciccarino}}},
  \bibinfo{author}{\bibfnamefont{R.}~\bibnamefont{{Bianco}}},
  \bibinfo{author}{\bibfnamefont{I.}~\bibnamefont{{Errea}}},
  \bibinfo{author}{\bibfnamefont{P.}~\bibnamefont{{Narang}}}, \bibnamefont{and}
  \bibinfo{author}{\bibfnamefont{B.~A.} \bibnamefont{{Bernevig}}},
  \bibinfo{journal}{arXiv e-prints} \bibinfo{eid}{arXiv:2305.02340}
  (\bibinfo{year}{2023}), \eprint{2305.02340}.

\bibitem[{\citenamefont{Parameswaran et~al.}(2013)\citenamefont{Parameswaran,
  Roy, and Sondhi}}]{Parameswaran2013}
\bibinfo{author}{\bibfnamefont{S.~A.} \bibnamefont{Parameswaran}},
  \bibinfo{author}{\bibfnamefont{R.}~\bibnamefont{Roy}}, \bibnamefont{and}
  \bibinfo{author}{\bibfnamefont{S.~L.} \bibnamefont{Sondhi}},
  \bibinfo{journal}{C.R. Phys.} \textbf{\bibinfo{volume}{14}},
  \bibinfo{pages}{816} (\bibinfo{year}{2013}).

\bibitem[{\citenamefont{Neupert et~al.}(2015)\citenamefont{Neupert, Chamon,
  Iadecola, Santos, and Mudry}}]{Neupert2015}
\bibinfo{author}{\bibfnamefont{T.}~\bibnamefont{Neupert}},
  \bibinfo{author}{\bibfnamefont{C.}~\bibnamefont{Chamon}},
  \bibinfo{author}{\bibfnamefont{T.}~\bibnamefont{Iadecola}},
  \bibinfo{author}{\bibfnamefont{L.}~\bibnamefont{Santos}}, \bibnamefont{and}
  \bibinfo{author}{\bibfnamefont{C.}~\bibnamefont{Mudry}},
  \bibinfo{journal}{Phys. Scr.} \textbf{\bibinfo{volume}{2015}},
  \bibinfo{pages}{014005} (\bibinfo{year}{2015}).

\bibitem[{\citenamefont{Mera and Ozawa}(2021{\natexlab{a}})}]{BMera2021}
\bibinfo{author}{\bibfnamefont{B.}~\bibnamefont{Mera}} \bibnamefont{and}
  \bibinfo{author}{\bibfnamefont{T.}~\bibnamefont{Ozawa}},
  \bibinfo{journal}{Phys. Rev. B} \textbf{\bibinfo{volume}{104}},
  \bibinfo{pages}{045104} (\bibinfo{year}{2021}{\natexlab{a}}).

\bibitem[{\citenamefont{Mera and Ozawa}(2021{\natexlab{b}})}]{BMera20212}
\bibinfo{author}{\bibfnamefont{B.}~\bibnamefont{Mera}} \bibnamefont{and}
  \bibinfo{author}{\bibfnamefont{T.}~\bibnamefont{Ozawa}},
  \bibinfo{journal}{Phys. Rev. B} \textbf{\bibinfo{volume}{104}},
  \bibinfo{pages}{115160} (\bibinfo{year}{2021}{\natexlab{b}}).

\bibitem[{\citenamefont{Carollo et~al.}(2020)\citenamefont{Carollo, Valenti,
  and Spagnolo}}]{CAROLLO2020}
\bibinfo{author}{\bibfnamefont{A.}~\bibnamefont{Carollo}},
  \bibinfo{author}{\bibfnamefont{D.}~\bibnamefont{Valenti}}, \bibnamefont{and}
  \bibinfo{author}{\bibfnamefont{B.}~\bibnamefont{Spagnolo}},
  \bibinfo{journal}{Physics Reports} \textbf{\bibinfo{volume}{838}},
  \bibinfo{pages}{1} (\bibinfo{year}{2020}).

\bibitem[{\citenamefont{Barthel and Schollw\"ock}(2008)}]{Barthel2008}
\bibinfo{author}{\bibfnamefont{T.}~\bibnamefont{Barthel}} \bibnamefont{and}
  \bibinfo{author}{\bibfnamefont{U.}~\bibnamefont{Schollw\"ock}},
  \bibinfo{journal}{Phys. Rev. Lett.} \textbf{\bibinfo{volume}{100}},
  \bibinfo{pages}{100601} (\bibinfo{year}{2008}).

\bibitem[{\citenamefont{Calabrese et~al.}(2011)\citenamefont{Calabrese, Essler,
  and Fagotti}}]{Calabrese2011}
\bibinfo{author}{\bibfnamefont{P.}~\bibnamefont{Calabrese}},
  \bibinfo{author}{\bibfnamefont{F.~H.~L.} \bibnamefont{Essler}},
  \bibnamefont{and} \bibinfo{author}{\bibfnamefont{M.}~\bibnamefont{Fagotti}},
  \bibinfo{journal}{Phys. Rev. Lett.} \textbf{\bibinfo{volume}{106}},
  \bibinfo{pages}{227203} (\bibinfo{year}{2011}).

\bibitem[{\citenamefont{Mitra}(2018)}]{Mitra2018}
\bibinfo{author}{\bibfnamefont{A.}~\bibnamefont{Mitra}},
  \bibinfo{journal}{Annual Review of Condensed Matter Physics}
  \textbf{\bibinfo{volume}{9}}, \bibinfo{pages}{245} (\bibinfo{year}{2018}).

\bibitem[{\citenamefont{Ozawa and Mera}(2021)}]{Ozawa20210}
\bibinfo{author}{\bibfnamefont{T.}~\bibnamefont{Ozawa}} \bibnamefont{and}
  \bibinfo{author}{\bibfnamefont{B.}~\bibnamefont{Mera}},
  \bibinfo{journal}{Phys. Rev. B} \textbf{\bibinfo{volume}{104}},
  \bibinfo{pages}{045103} (\bibinfo{year}{2021}).

\bibitem[{Sup()}]{Suppl}
\bibinfo{howpublished}{See the supplementary material for the details of the
  exact proof as well as the numerical verification of our results for other
  models}.

\bibitem[{\citenamefont{Kitaev}(2001)}]{Kitaev2001}
\bibinfo{author}{\bibfnamefont{A.}~\bibnamefont{Kitaev}},
  \bibinfo{journal}{Phys.-Usp.} \textbf{\bibinfo{volume}{44}},
  \bibinfo{pages}{131} (\bibinfo{year}{2001}).

\bibitem[{\citenamefont{Su et~al.}(1979)\citenamefont{Su, Schrieffer, and
  Heeger}}]{Su1979}
\bibinfo{author}{\bibfnamefont{W.~P.} \bibnamefont{Su}},
  \bibinfo{author}{\bibfnamefont{J.~R.} \bibnamefont{Schrieffer}},
  \bibnamefont{and} \bibinfo{author}{\bibfnamefont{A.~J.}
  \bibnamefont{Heeger}}, \bibinfo{journal}{Phys. Rev. Lett.}
  \textbf{\bibinfo{volume}{42}}, \bibinfo{pages}{1698} (\bibinfo{year}{1979}).

\bibitem[{\citenamefont{Qi et~al.}(2006)\citenamefont{Qi, Wu, and
  Zhang}}]{Qi2006}
\bibinfo{author}{\bibfnamefont{X.-L.} \bibnamefont{Qi}},
  \bibinfo{author}{\bibfnamefont{Y.-S.} \bibnamefont{Wu}}, \bibnamefont{and}
  \bibinfo{author}{\bibfnamefont{S.-C.} \bibnamefont{Zhang}},
  \bibinfo{journal}{Phys. Rev. B} \textbf{\bibinfo{volume}{74}},
  \bibinfo{pages}{085308} (\bibinfo{year}{2006}).

\bibitem[{\citenamefont{Cai et~al.}(2008)\citenamefont{Cai, Chen, Kou, and
  Wang}}]{Cai2008}
\bibinfo{author}{\bibfnamefont{Z.}~\bibnamefont{Cai}},
  \bibinfo{author}{\bibfnamefont{S.}~\bibnamefont{Chen}},
  \bibinfo{author}{\bibfnamefont{S.}~\bibnamefont{Kou}}, \bibnamefont{and}
  \bibinfo{author}{\bibfnamefont{Y.}~\bibnamefont{Wang}},
  \bibinfo{journal}{Phys. Rev. B} \textbf{\bibinfo{volume}{78}},
  \bibinfo{pages}{035123} (\bibinfo{year}{2008}).

\bibitem[{\citenamefont{Souza et~al.}(2000)\citenamefont{Souza, Wilkens, and
  Martin}}]{Souza2000}
\bibinfo{author}{\bibfnamefont{I.}~\bibnamefont{Souza}},
  \bibinfo{author}{\bibfnamefont{T.}~\bibnamefont{Wilkens}}, \bibnamefont{and}
  \bibinfo{author}{\bibfnamefont{R.~M.} \bibnamefont{Martin}},
  \bibinfo{journal}{Phys. Rev. B} \textbf{\bibinfo{volume}{62}},
  \bibinfo{pages}{1666} (\bibinfo{year}{2000}).

\bibitem[{\citenamefont{Francesco et~al.}(2012)\citenamefont{Francesco,
  Mathieu, and Senechal}}]{Francesco:2012aa}
\bibinfo{author}{\bibfnamefont{P.}~\bibnamefont{Francesco}},
  \bibinfo{author}{\bibfnamefont{P.}~\bibnamefont{Mathieu}}, \bibnamefont{and}
  \bibinfo{author}{\bibfnamefont{D.}~\bibnamefont{Senechal}},
  \emph{\bibinfo{title}{Conformal field theory}} (\bibinfo{publisher}{Springer
  Science and Business Media}, \bibinfo{year}{2012}).

\bibitem[{\citenamefont{O.Wehling et~al.}(2014)\citenamefont{O.Wehling,
  Black-Schaffer, and Balatsky}}]{Wehling:2014aa}
\bibinfo{author}{\bibfnamefont{T.}~\bibnamefont{O.Wehling}},
  \bibinfo{author}{\bibfnamefont{A.~M.} \bibnamefont{Black-Schaffer}},
  \bibnamefont{and} \bibinfo{author}{\bibfnamefont{A.~V.}
  \bibnamefont{Balatsky}}, \bibinfo{journal}{Advances in Physics}
  \textbf{\bibinfo{volume}{63}}, \bibinfo{pages}{1} (\bibinfo{year}{2014}).

\bibitem[{\citenamefont{Gogolin et~al.}(2004)\citenamefont{Gogolin, Nersesyan,
  and Tsvelik}}]{Gogolin:2004aa}
\bibinfo{author}{\bibfnamefont{A.~O.} \bibnamefont{Gogolin}},
  \bibinfo{author}{\bibfnamefont{A.~A.} \bibnamefont{Nersesyan}},
  \bibnamefont{and} \bibinfo{author}{\bibfnamefont{A.~M.}
  \bibnamefont{Tsvelik}}, \emph{\bibinfo{title}{Bosonization and strongly
  correlated systems}} (\bibinfo{publisher}{Cambridge university press},
  \bibinfo{year}{2004}).

\end{thebibliography}

\end{document}